\documentclass[iop]{emulateapj}

\def\Msun{M_\odot}
\def\microas{\mu{\rm as}}

\begin{document}

\title{The PHASES Differential Astrometry Data Archive.  III.  Limits to Tertiary 
Companions}

\author{Matthew W.~Muterspaugh\altaffilmark{1, 2}, 
Benjamin F.~Lane\altaffilmark{3}, S.~R.~Kulkarni\altaffilmark{4}, 
Maciej Konacki\altaffilmark{5, 6}, Bernard F.~Burke\altaffilmark{7}, 
M.~M.~Colavita\altaffilmark{8}, M.~Shao\altaffilmark{8}}
\altaffiltext{1}{Department of Mathematics and Physics, College of Arts and 
Sciences, Tennessee State University, Boswell Science Hall, Nashville, TN 
37209}
\altaffiltext{2}{Tennessee State University, Center of Excellence in 
Information Systems, 3500 John A. Merritt Blvd., Box No.~9501, Nashville, TN 
37209-1561}
\altaffiltext{3}{Draper Laboratory,  555 Technology Square, Cambridge, MA 
02139-3563}
\altaffiltext{4}{Division of Physics, Mathematics and Astronomy, 105-24, 
California Institute of Technology, Pasadena, CA 91125}
\altaffiltext{5}{Nicolaus Copernicus Astronomical Center, Polish Academy of 
Sciences, Rabianska 8, 87-100 Torun, Poland}
\altaffiltext{6}{Astronomical Observatory, Adam Mickiewicz University, 
ul.~Sloneczna 36, 60-286 Poznan, Poland}
\altaffiltext{7}{MIT Kavli Institute for Astrophysics and Space Research, 
MIT Department of Physics, 70 Vassar Street, Cambridge, MA 02139}
\altaffiltext{8}{Jet Propulsion Laboratory, California Institute of 
Technology, 4800 Oak Grove Dr., Pasadena, CA 91109}

\email{matthew1@coe.tsuniv.edu, blane@draper.com, maciej@ncac.torun.pl}

\begin{abstract}
The Palomar High-precision Astrometric Search for Exoplanet Systems (PHASES) 
monitored 51 subarcsecond binary systems to evaluate whether tertiary 
companions as small as Jovian planets orbited either the primary or secondary 
stars, perturbing their otherwise smooth Keplerian motions.  
Twenty-one of those 
systems were observed 10 or more times and show no evidence of additional 
companions.  A new algorithm is presented for identifying astrometric 
companions and establishing the (companion mass)-(orbital period) 
combinations that 
can be excluded from existence with high confidence based on the PHASES 
observations, and the regions of mass-period phase space being excluded are 
presented for 21 PHASES binaries.
\end{abstract}


\section{Introduction}

Searches for planets in close binary systems explore the degree to which 
stellar multiplicity inhibits or promotes planet formation.  The orbits in 
which planets in binary systems can be stable are divided into three 
classes:  (1) 
P-type (for ``Planetary Type''), or circumbinary planets, which orbit both 
stars at a separation much larger than that of the stars themselves, 
(2) S-type (for ``Satellite Type'') which orbit either the primary or the 
secondary star but not both, with an orbital size much smaller than the 
distance between the stars, and (3) L-type, for planets found at Lagrangian 
points \citep{Dvorak1982}.  The Palomar High-Precision Astrometric Search for 
Exoplanet Systems (PHASES) was a search at the Palomar Testbed 
Interferometer \citep[PTI;][]{col99} targeting 51 close binaries (semimajor 
axis a few 100 milliarcseconds) to identify S-type planetary companions to 
either star in each pair by measuring the relative separations of the stars 
with $\sim 35 \microas$ astrometric precisions \citep{LaneMute2004a}.

Current theory is that planets form in and from material of dusty disks 
observed around young stars.  Some models in which giant planet formation 
occurs over large amounts of time 
(e.g., the core-accretion scenario at 1-10 Myr) predict that an 
extra-turbulent environment, such as those around binary stars, will disrupt 
planet formation by dispersing the protoplanetary disk while it is young or 
increasing impact speeds between planetesimals preventing accretion into 
larger objects \citep{Mar:00::, Mar:07}.  
If the timescale is short (as in the gravitational instability theory), the 
process can happen before the disk is disrupted, 
or even be enhanced due to additional instabilities in 
the planet forming disks \citep{Boss98}.  

The leading theories have helped promote a common belief that 
planet formation is difficult or inhibited in binary or multiple stars 
because these disks might be more short-lived.  However, a lower limit 
of $22\%$ of known planet hosting stars have distant stellar companions 
\citep{Raghavan2005}.  Given that multiplicity is the norm in the solar 
neighborhood \citep[$57\%$;][]{Duq:91} and star-forming regions 
\citep{Sim:95::}, the entire issue of planets in binary and multiple stars 
cannot be ignored if a complete census of planets is to be taken.  
Indeed, searching for planets in such systems acts as a 
test of planet formation models in complex dynamical environments 
\cite[see, for example,][]{desidera2006}.

While planets in binaries appear to be common, most of the binaries being 
surveyed have very wide separations and the companion star has little 
gravitational influence on the environment of the planet host.  PHASES was 
different in that the stellar companions were much closer to the planet-hosting 
star---only a handful of binaries targeted by other programs have these 
small physical separations.  These systems place much stronger constraints on 
the impact of dynamics on planet formation.

Studying relatively close pairs of stars, where dynamic perturbations 
are the strongest, provides the most restrictive constraints of this type 
\citep[see, for example,][]{thebault2004}.  Searching for planets in those 
systems can determine whether the planet formation mechanisms found in nature 
are sensitive to binary dynamics or not, a property which must be matched by 
theoretical models \citep{hatzes2005}.  
It may be that multiple mechanisms contribute to 
giant planet formation in nature.  Establishing the rate at which giant 
planets exist in binaries will distinguish the relative frequencies at which 
different processes contribute.

It can be shown that dynamic interactions between stars in young clusters can 
result in close binaries ($a < 50$ AU) having S-type planetary companions that 
did not form in situ in the close binary, but around a single star, which 
later interacted with a binary, inserting the planet into the system 
\citep{Pfahl2005}.  The low frequency of these interactions would 
result in less than $0.1\%$ of such binaries hosting planets were this the 
only mechanism from which such configurations arise \citep{PfahlMute2006}.  Any 
planet frequency above this level would indicate that the planet formation 
process can survive the binary star environment---the number of planets beyond 
that frequency must have formed in situ.

A few close binaries have been identified hosting giant planets and are 
listed in Table \ref{tab:tabclose}.  The 5-6 such 
systems already identified represent a larger frequency of 
occurrence than such dynamics could explain.  Ongoing efforts to identify 
such systems will need to concentrate on better identifying the statistics of 
the total number of close binaries that have been included in surveys in order 
to better understand the planet frequency statistics.  In this paper, the null 
results for the PHASES effort are reported to quantify the population 
statistics of this search for comparison with the number of candidates 
discovered (see Paper V).

\begin{deluxetable}{llcccc}
\tablecolumns{6}
\tablecaption{Close Binaries with Planets. \label{tab:tabclose}}
\tablehead{
\colhead{Object} &
\colhead{$a$(AU)} &
\colhead{$e$\tablenotemark{a}} &
\colhead{$M_1/M_2$\tablenotemark{b}} &
\colhead{$R_t$(AU)\tablenotemark{c}} &
\colhead{References}
}
\startdata
HD 188753\tablenotemark{d} & 12.3     & 0.50    & 1.06/1.63 & 1.3     & 1, 2, 3 \\
$\gamma$ Cephei            & 18.5     & 0.36    & 1.59/0.34 & 3.6     & 4, 5 \\
GJ 86 \tablenotemark{e}    & $\sim$20 & \nodata & 0.7/1.0   & $\sim$5 & 6, 7 ,8 \\
HD 41004\tablenotemark{f}  & $\sim$20 & \nodata & 0.7/0.4   & $\sim$6 & 9 \\
HD 126614                  & $\sim$45 & \nodata & 1.145/0.324 &$\sim$15 & 10 \\
HD 196885                  & $\sim$25 & \nodata & 1.3/0.6   & $\sim$8 & 11 
\enddata
\tablenotetext{a}{When the eccentricity is unknown, the projected binary separation is used as an approximation, except 
in the case of HD 126614, where a linear velocity trend due to the star is observed, and the binary itself has been 
resolved, leading to two possible solutions with $a = 40^{+7}_{-4}$ and $50^{+2}_{-3}$ AU.}
\tablenotetext{b}{Mass of star hosting planet divided by mass of the companion star.}
\tablenotetext{c}{The distance from the primary star at which a disk would be rapidly 
truncated by tides \citep{Pichardo2005}.}
\tablenotetext{d}{The companion star itself is a binary with the semimajor axis 0.67\,AU.  
This candidate is controversial due to minimal data in the discovery paper with sporadic 
observing cadence and a lack of evidence found by \cite{2007A&A...466.1179E} and \cite{2009MNRAS.399..906M}.}
\tablenotetext{e}{The companion star is a white dwarf of mass $\simeq$$0.5\Msun$.  
To estimate $R_t$ at the time of formation, an original companion mass of $1\Msun$ is assumed.}
\tablenotetext{f}{The secondary also has a substellar companion--a brown dwarf with a 1.3\,day period.}
\tablerefs{
(1) Konacki 2005; \nocite{Konacki2005}
(2) Eggenberger et al.~2007; \nocite{2007A&A...466.1179E}
(3) Mazeh et al.~2009; \nocite{2009MNRAS.399..906M}
(4) Campbell et al.~1988; \nocite{Campbell1988}
(5) Hatzes et al.~2003; \nocite{Hatzes2003}
(6) Queloz et al.~2000; \nocite{Queloz2000} 
(7) Mugrauer \& Neuh{\"a}user 2005; \nocite{Mugrauer2005} 
(8) Lagrange et al.~2006; \nocite{Lagrange2006}
(9) Zucker et al.~2004; \nocite{Zuc2004} 
(10) Howard et al.~2010 \nocite{howard2010}
(11) Chauvin et al.~2006 \nocite{Chauvin2006}
}
\end{deluxetable}

This paper is the third in a series analyzing the final results of the PHASES 
project as of its completion in late 2008.  The first paper describes the 
observing method, sources of measurement uncertainties, limits of observing 
precisions, derives empirical scaling rules to account for noise sources 
beyond those predicted by the standard reduction algorithms, and presents the 
full catalog of astrometric measurements from PHASES \citep{Mute2010A}.  The 
second paper combines PHASES astrometry with astrometric measurements made by 
other methods as well as radial velocity observations (where available) to 
determine orbital solutions to the binaries' Keplerian motions, determining 
physical properties such as component masses and system distance when 
possible \citep{Mute2010B}.  The current paper 
presents limits on the existence of substellar tertiary companions orbiting 
either the primary or secondary stars in those systems that are found to be 
consistent with being simple binaries.  Paper IV presents 
orbital solutions to a known triple star system (63 Gem $=$ HD 58728) and a 
newly discovered triple system (HR 2896 $=$ HD 60318) \citep{Mute2010D}.  
Finally, Paper V 
presents candidate substellar companions to PHASES binaries as detected by 
astrometry \citep{Mute2010E}.

Astrometric measurements were made 
as part of the PHASES program at PTI, which was 
located on Palomar Mountain near San Diego, CA.  It was developed by the Jet 
Propulsion Laboratory, California Institute of Technology for NASA, as a 
testbed for interferometric techniques applicable to the Keck Interferometer 
and other missions such as the Space Interferometry Mission (SIM).  It 
operated in the J ($1.2 \mu{\rm m}$), H ($1.6 \mu{\rm m}$), and K 
($2.2 \mu{\rm m}$) bands, and combined starlight from two out of three 
available 40-cm apertures.  The apertures formed a triangle with one 110 and 
two 87 meter baselines.  PHASES observations began in 2002 continued through 
2008 November when PTI ceased routine operations.

\section{Companion Search Algorithm}

PHASES differential astrometry measurements are presented in Paper I.  This 
includes corrections to the measurement uncertainties, which are used here.  

\cite{Mute06Limits} presented an initial analysis algorithm and preliminary 
results for the range of mass-period phase space in which tertiary companions 
can be ruled out for eight binaries.  This initial algorithm had some 
limitations that have since been improved upon.  These limitations included 
the fact that only face-on, circular companion orbits were modeled, 
the algorithm used a statistical analysis that did not account 
for how the observing cadence can impact the false alarm rate, 
and the algorithm was very computationally intensive.
An alternative analysis method has been developed for identifying candidate 
companions and establishing the range of mass-period pairings for hypothetical 
tertiary companions that can be ruled out by the PHASES observations.  While 
still relatively computationally intensive, 
the new algorithm is less so and solves 
the other limitations much more completely.

\subsection{Identifying Candidate Objects}

\cite{cumming1999} developed a general method for (1) identifying Keplerian 
signals, (2) estimating the level of confidence in the signal detection, and 
(3) evaluating the mass threshold (as a function of orbital period) that can 
be shown not to exist by a given data set, to some level of confidence.  This 
algorithm has been modified for use with the PHASES measurements.  The major 
differences arise from PHASES being astrometric measurements (whereas 
\cite{cumming1999} analyzed velocity measurements), the PHASES measurements 
were two-dimensional in nature, and because the stars are binary and in orbit 
around each other some degree of model-fitting is necessary even in the case 
that additional companions are not found.  In other words, the no-companion 
model is not a constant value, but rather the orbit of the binary itself.  The 
model with an additional companion is the binary orbit plus the companion 
orbit.  When evaluating the model that includes an additional companion, 
it is also crucial to reoptimize 
the parameters associated with the binary orbit itself, to adjust for the 
addition of the perturbation orbit.  In other words, it is not enough to fit 
a companion model to residuals that were computed by subtracting an optimized 
binary model from the original measurements---both components of the model need 
to be reoptimized.  These modifications to the work of \cite{cumming1999}, 
and the code base for it, were designed, developed, and tested by the first 
author's (M.W.M.) team during the SIM 
Double Blind Test \citep{2009AAS...21330001T, 2010EAS....42..191T}, 
and were demonstrated as reliable during that time.

First, a single Keplerian orbital model was fit to the PHASES data for each 
star, and the parameters of that orbit were optimized to minimize the $\chi^2$ 
goodness-of-fit metric.  These served as comparison models against which to 
compare how well the data were represented by alternative models, such as those 
with the Keplerian plus a perturbation caused by the reflex motion of one star 
as an additional object orbits it.  
The best fit $\chi^2$ of the Keplerian orbital 
model is $\chi^2_2$, where the 2 subscript indicates only two 
objects are in the system (the stars of the binary itself).

Second, a double Keplerian orbital model was fit to the data at several 
possible values of the companion orbital period.  During this fitting, all 
parameters of the known binary orbit were reoptimized (being seeded at their 
values from the best-fit single Keplerian model to initiate the fitting), as 
well as the orbital elements of the perturbation model, excluding eccentricity 
(which was set to zero), time of periastron passage $T_\circ$ (because it would 
be degenerate with the Campbell parameter $\omega$ when the eccentricity is 
zero), and orbital period (which was fixed at its seed value, as described 
below).  Thus, while the binary orbit was a full Keplerian model, the 
perturbation model was circular only, though at any inclination and orientation 
on the sky.  However, in practice the circular orbit model 
correctly identified most companions having eccentric orbits as well.  The 
orbital model was optimized in the Thiele-Innes parameter set 
rather than the normal Campbell parameter set to improve computational 
efficiency \citep[for a recent review, see][]{wrightHoward2009}.  A downhill 
search algorithm was used to minimize $\chi^2$ of the model in a variant of the 
standard Levinberg-Marquart approach.

The companion orbital periods at which the double Keplerian model was 
evaluated were selected in a method inspired by Nyquist frequency sampling.  
For a data set spanning time $T$, the set of periods selected was given 
by $P = 2 f T / k$, 
where $k$ is a positive integer, and $f$ is an oversampling factor.  If the 
times of the data measurements were uniform, $f=1$ could be safely assumed; 
however, this is not the case for real measurements.  Thus, $f=3$ was chosen as 
the oversampling factor in the present analysis to ensure sampling density 
did not cause potential companions to be missed.  The largest value of $k$ was 
chosen to be that for which $P=6$ days, both for computational efficiency, and 
because astrometry is unlikely to find many objects at shorter orbital periods 
that are not already known from radial velocity measurements.  It was 
important to examine periods this short to explore effects at the 
$\sim {\rm week}$ cadence common for the PHASES observations.  Finally, 
because some orbital curvature could be observed for massive companions with 
orbital periods longer than the data span $T$, one additional value of $k=1/2$ 
was also evaluated (making the longest period evaluated $P = 2 f T/k = 12 T$).

At each value of the perturbation orbit's period, the best-fitting model's 
value of $\chi^2_3$ is evaluated (here, the subscript 3 indicates the model 
represents three objects are in the system).  These were used to create a 
periodogram similar to those in \cite{cumming1999}, for which the largest peak 
corresponds to the value that best improved the fit to the data.  The 
periodogram values were calculated as
\begin{equation}
z(P) = \left(\frac{2N-11}{11-7}\right)\left( \frac{\chi^2_2 - \chi^2_3(P)}{\chi^2_3(P_{\rm best})}\right)
\end{equation}
where $N$ is the number of two-dimensional astrometric measurements (thus, 
$2N$ total measurements were analyzed), 11 is the number of free parameters 
in the 
double Keplerian model (the normal seven Keplerian parameters for the binary 
orbit, and only four for the perturbation orbit were free parameters, since 
the epoch of periastron passage $T_\circ$, the eccentricity $e=0$, and orbital 
period $P$ were held fixed during model fitting), 7 is the number of free 
parameters in the single Keplerian model, and $\chi^2_3(P_{\rm best})$ is the 
overall best (smallest) value of $\chi^2_3$ of all the periods sampled.  This 
statistic follows the $F$ distribution as a test of whether the addition of the 
second orbit is valid \citep[see, for example,][]{bevingtonRobinson}.

\subsection{False Alarm Probability of Companion Detection}

Because sampling cadence can have effects on the periodogram that are not 
straightforward to calculate, the false alarm probability (FAP) 
of a given value of 
$z$ was calculated by creating synthetic data sets with identical cadence and 
scatter in the data, rather than directly from the expected $F$ distribution.  
\cite{cumming1999} identify two ways of creating synthetic data sets, and note 
that in practice, the two approaches produce very similar results.  One 
approach is to scramble residuals from the actual measurements, rescaling their 
values by the ratio of the uncertainties of the replacement measurement and the 
one actually made at the given time.  This has the advantage that the synthetic 
data set has similar statistical properties to the actual data and does not 
assume Gaussian (or other) statistics to the data.  The alternative is to 
create synthetic data sets from a random number generator, scaling the random 
numbers by the measurement uncertainty of a given measurement.  
The first approach was 
made more difficult for the PHASES measurements, given their two-dimensional 
nature, variable measurement uncertainties, variable orientations of the error 
ellipses on the sky, and the presence of the motion of the binary as a whole.  
Thus, for the present analysis, the latter approach was selected.  A random 
number generator created a list of Gaussian-distributed random numbers.  The 
use of Gaussian statistics for the synthetic noise was justified by the 
distribution of the residuals from PHASES measurements, as demonstrated in 
Paper I.  For 
each measurement, two random numbers were used to create synthetic data 
in the basis of the measurement's uncertainty ellipse minor and major axis 
(in this basis, the two-dimensional 
uncertainties have zero covariance) and those 
values are then rotated into the right ascension-declination basis in 
accordance with the uncertainty covariance.  The 
best-fit single-Keplerian signal was then added to the random values, creating 
a complete synthetic data set representing the binary motion, but no additional 
real perturbations.  The synthetic data set was analyzed in the same manner as 
the real data set, and the maximum value of $z$ for that synthetic 
data was recorded.  The process was repeated 1000 times, each time creating a 
new synthetic data set.  The fraction of 
synthetic data sets producing maximum values of $z$ greater than that observed 
in the actual data determines the level of confidence, or FAP, 
that the peak in the data periodogram represents a 
perturbation created by a real object, rather than being a statistical 
fluctuation.  The value of $z$ of the tenth largest maximum values of $z$ from 
the synthetic data identifies the level at which a detected signal would have 
an FAP of $10/1000 = 1\%$.

\subsection{Detection Limits}

After computing the periodogram and the FAP of its values, the tertiary 
companion masses which can be shown {\em not} to exist with high confidence 
were evaluated, as a function of the orbital period of the tertiary 
companion.  This 
represents the sensitivity limits of the PHASES survey.  For each orbital 
period $P$ for a potential perturber, 1000 synthetic data sets were produced as 
above, but both the binary plus 
an additional Keplerian signal were added to the data set 
representing a tertiary companion to the system.  The parameters describing 
the second Keplerian were selected in the Campbell set as follows.  
\begin{itemize}
\item the orbital period $P$ was given by the tertiary companion orbital period 
being evaluated,
\item the epoch of periastron passage $T_\circ$ was selected from a flat 
distribution centered at the average time of observation, and covering a span 
equal to the orbital period $P$ (this range covers all possible 
non-degenerate values of $T_\circ$), 
\item the eccentricity was selected from a flat distribution between 
$0 \le e \le 0.5$; in practice, the results were fairly accurate for any value 
of eccentricity, 
\item the inclination was selected from a flat distribution in $\sin i$, 
\item $\omega$ was selected from a flat distribution between 
$0^\circ \le \omega \le 360^\circ$, 
\item $\Omega$ was selected from a flat distribution between 
$0^\circ \le \Omega \le 360^\circ$, and 
\item the semi-major axis $a$ was given an initial value close to the average 
minor axis uncertainty of the PHASES measurements, though this will be modified 
upon iteration, as described below.
\end{itemize}
For the elements that are chosen randomly ($T_\circ$, $e$, $i$, $\omega$, and 
$\Omega$), different values were selected for each of the 1000 synthetic data 
sets being created.  Each of the 1000 synthetic data sets were fit to the 
double Keplerian model; the fit was seeded with the known orbital parameters 
(with the exception of the perturbing orbit's eccentricity, which was fixed at 
zero regardless of the actual eccentricity used to generate the synthetic 
data for equality with the actual search algorithm on real data).  
During fitting, all seven parameters of the binary orbit were free 
parameters, as well as $i$, $\omega$, $\Omega$, and $a$ of the tertiary 
companion orbit.  Also, the best-fit single Keplerian model for the synthetic 
data set was computed, for use in evaluating $z$, as
\begin{equation}
z(P) = \left(\frac{2N-11}{11-7}\right)\left( \frac{\chi^2_2 - \chi^2_3(P)}{\chi^2_3(P)}\right)
\end{equation}
where $\chi^2_3(P_{\rm best})$ was replaced by $\chi^2_3$ in the denominator, 
since only one orbital period was being evaluated.  The fraction of synthetic 
data sets with $z$ exceeding the maximum value of $z$ in the actual data (of 
all orbital periods evaluated) was computed.  If the fraction was larger than 
some specified confidence level (here, 99\%), the semimajor axis of the 
perturbing orbit was decreased for the next iteration; if it was smaller, the 
semimajor axis was increased.  This procedure was iterated, each time 
generating 
1000 new synthetic data sets, until the semimajor axis that creates synthetic 
data sets for which 99\% were found to have $z$ exceeding that of the data was 
bounded.  Once bounded, further iterations refined this bound until the correct 
semimajor axis was determined to a precision of $4 \, {\rm \mu as}$ or better 
(corresponding to roughly 1/10 the typical minor axis uncertainty of PHASES 
measurements).  The resulting limiting semimajor axis was converted into the 
corresponding companion mass necessary to create a reflex motion of one of the 
stars in the binary by that amount, given the star's mass, the tertiary 
companion orbital period, and the overall distance to the star system.

\subsection{Stability of Orbits}

Finally, there is the question of whether the orbits are stable, since the 
presence of the second star creates a different dynamical environment.  Indeed, 
this is part of the motivation for searching for planets in binaries separated 
by only 10-50 AU:  whether the formation mechanism for giant planets can 
survive such a dynamic environment.  System stability offers an external check 
for whether a candidate companion is a false identification.  The empirical 
stability rules identified by \cite{holman1999} are calculated for each binary 
and set as approximate limits for the ranges over which companions might be 
expected to have stable orbits using the following relationship:
\begin{equation}
a_c = (0.464 -0.380 \mu - 0.631 e  + 0.586 \mu e + 0.150 e^2 - 0.198 \mu e^2) a_b
\end{equation}
where $a_c$ is the semimajor axis of the largest stable orbit, $a_b$ is the 
semimajor axis of the binary, $e$ is the eccentricity of the binary, and 
$\mu = m_2/(m_1+m_2)$ is the mass ratio of the binary, where $m_2$ is the perturbing 
star and $m_1$ hosts the tertiary companion.  The limiting orbital periods are listed 
in Table \ref{tab::stable}.

\begin{deluxetable}{lrrrr}
\tablecolumns{5}
\tablewidth{0pc} 
\tablecaption{Maximum Stable Orbital Periods and Star Masses and Distances Used To Compute Limits\label{tab::stable}}
\tablehead{ 
\colhead{HD Number} & \colhead{$P_{1, \, {\rm max}}$} & \colhead{$P_{2, \, {\rm max}}$} & \colhead{${M_{\rm star} \, (\Msun)}$} & \colhead{${d_{\rm star}} \, ({\rm pc})$}}
\startdata
5286    & 6354 & 6354 & 1.00 & 38.92 \\
6811    & 16804 & 16804 & 3.55 & 225.73 \\
17904   &  170 & 170 & 2.06 & 72.10 \\
26690   &  284 & 231 & 0.82 & 36.48 \\
44926   & 32410 & 32410 & 6.00 & 438.60 \\
76943   & 1284 & 1124 & 1.04 & 16.26 \\
77327   &  591 & 591 & 5.20 & 129.70 \\
81858   & 2450 & 1663 & 1.10 & 36.36 \\ 
114378  &  552 & 552 & 1.22 & 17.89 \\ 
114378  &  555 & 544 & 1.22 & 17.89 \\ 
129246  &  N/A & N/A & 66.00 & 55.34 \\
137107  & 1738 & 1646 & 1.10 & 18.50 \\
137391  &  155 & 155 & 1.62 & 36.10 \\
137909  &  197 & 181 & 1.33 & 34.12 \\
140159  & 1357 & 1357 & 1.86 & 57.80 \\
140436  & 2087 & 2087 & 1.86 & 43.29 \\
155103  &  149 & 149 & 1.66 & 55.56 \\
187362  &   87 &  87 & 2.35 & 100.10 \\
202275  &  148 & 148 & 1.19 & 18.38 \\ 
202275  &  157 & 153 & 1.19 & 18.38 \\ 
202444  & 2221 & 2221 & 1.31 & 20.37 \\
207652  & 1209 & 1209 & 1.32 & 33.78 \\
214850  &  130 & 133 & 1.07 & 34.43 
\enddata
\tablecomments{
The maximum stable orbital periods for 
tertiary companions to the 21 binaries under 
consideration, and the values of stellar 
masses and system distances from the Sun used 
to convert astrometric perturbation amplitude 
to companion mass.  Column 1 is the binary's 
HD number.  Columns 2 and 3 are the maximum stable 
orbital periods in days for S-type planets, calculated 
according to the formula by \cite{holman1999}.  
Columns 4 and 5 are the stellar mass and distance 
to the system, respectively; when only visual orbits were 
available, the mass used is that of the average 
component, whereas for systems having radial velocity 
measurements, the lower mass component is assumed, except 
in the case of HD 81858, for which the mass ratio has 
large uncertainty and the average component mass is 
used.  Two entries are present for HD 114378 and HD 202275; 
these systems were specifically modeled by \cite{holman1999}.  
The first entry lists the maximum stable orbital periods 
according to their formula, whereas the second entry lists 
the actual value they list in their Table 4.
}
\end{deluxetable}

\section{Companion Limits for Specific Systems}

In this section, the mass-period pairings for tertiary companions that can be 
ruled out for each of the 21 binaries are presented.  In Figures 
\ref{fig::5286_phase_space}--\ref{fig::214850_phase_space_all}, 
regions shaded gray 
indicate companion orbital periods that are not expected to be dynamically 
stable.

\subsection{HD 5286}

\begin{figure*}[!ht]
\plottwo{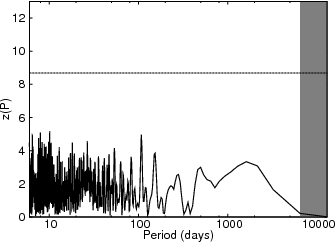}{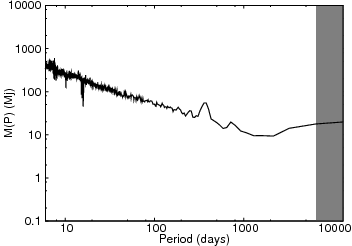}
\caption[HD 5286 ($\phi$ And) Mass-Period Companion Phase Space] 
{ \label{fig::5286_phase_space}
z-periodogram (left) and the mass-period companion phase space for HD 5286 (right).  
Companions in the regions above the plotted exclusion curve with circular 
orbits with any orientation are not consistent with the PHASES observations, 
with 99\% confidence.  Companions as small as 
9.3 Jupiter masses can be ruled out by PHASES observations.}
\end{figure*}

HD 5286 (36 And, HR 258, HIP 4288, WDS 00550$+$2338) is a pair of 
subgiant stars with spectral types G6 and K6.  The FAP of the highest peak 
($z = 5.19$ at $P=10.1$ days) in the z-periodogram is 23.0\%.  The 99\% 
confidence level would have been at $z = 8.68$.  
Figure \ref{fig::5286_phase_space} 
shows the periodogram and region of mass-period 
space in which companions can be 
ruled out with 99\% confidence.

\subsection{HD 6811}

\begin{figure*}[!ht]
\plottwo{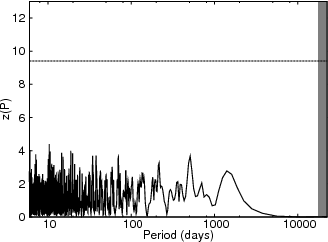}{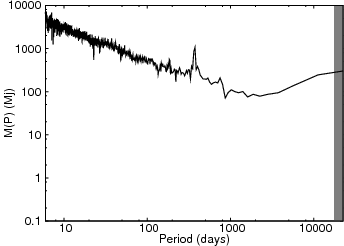}
\caption[HD 6811 ($\phi$ And) Mass-Period Companion Phase Space] 
{ \label{fig::6811_phase_space}
z-periodogram (left) and the mass-period companion phase space for HD 6811 (right).  
Companions in the regions above the plotted exclusion curve with circular 
orbits with any orientation are not consistent with the PHASES observations, 
with 99\% confidence.  Companions as small as 
71 Jupiter masses can be ruled out by PHASES observations.}
\end{figure*}

HD 6811 ($\phi$ And, 42 And, HR 335, HIP 5434, WDS 01095$+$4715) is 
a pair of massive, distant B stars (B6IV and B9V).  As a result, astrometry has 
more limited sensitivity to tertiary companions.  The FAP of the highest peak 
($z = 4.40$ at $P=10.4$ days) in the z-periodogram 
is 47.2\%.  The 99\% confidence level would have been $z = 9.40$.  
The periodogram and limits to tertiary companions are plotted in 
Figure \ref{fig::6811_phase_space}.  

\subsection{HD 17904}
\begin{figure*}[!ht]
\plottwo{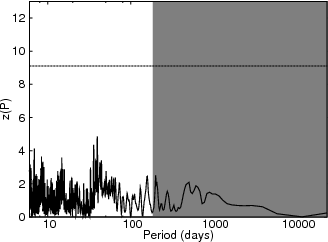}{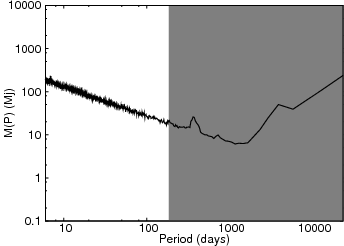}
\caption[HD 17904 (20 Per) Mass-Period Companion Phase Space] 
{ \label{fig::17904_phase_space}
z-periodogram (left) and the mass-period companion phase space for HD 17904 (right).  
Companions in the regions above the plotted exclusion curve with circular 
orbits with any orientation are not consistent with the PHASES observations, 
with 99\% confidence.  Companions as small as 
18 Jupiter masses can be ruled out by PHASES observations.}
\end{figure*}

The periodogram and limits to tertiary companions to HD 17904 
(20 Per, HR 855, HIP 13490, WDS 02537$+$3820) are plotted in Figure 
\ref{fig::17904_phase_space}.  The 1269 day subsystem suggested by \cite{abt1976} is 
not seen, nor would it be predicted to be stable if it did exist.  
This is consistent with the radial velocity studies by \cite{scarfe1978} 
and \cite{morbey1987} who also found no evidence of such a subsystem.  
The FAP of the highest peak ($z = 4.87$ at $P=39.4$ days) in the z-periodogram 
is 46.2\% and $z=9.10$ would be necessary to reach the 99\% confidence level.

\subsection{HD 26690}

\begin{figure*}[!ht]
\plottwo{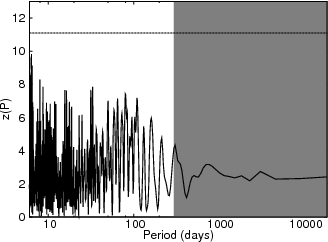}{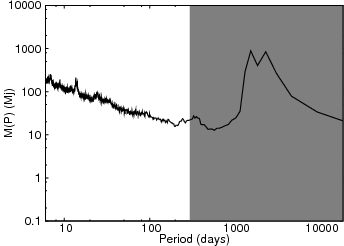}
\caption[HD 26690 (46 Tau) Mass-Period Companion Phase Space for PHASES Astrometry] 
{ \label{fig::26690_phase_space_phases}
z-periodogram (left) and the mass-period companion phase space for HD 26690 (right).  
Companions in the regions above the plotted exclusion curve with circular 
orbits with any orientation are not consistent with the PHASES observations, 
with 99\% confidence.  Companions as small as 
15.6 Jupiter masses in stable orbits can be ruled out by PHASES observations.}
\end{figure*}

\begin{figure*}[!ht]
\plottwo{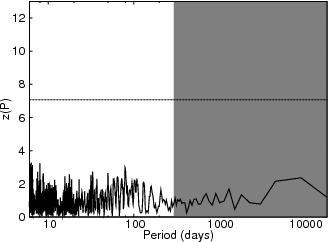}{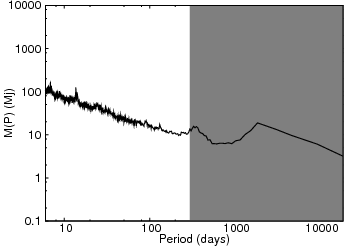}
\caption[HD 26690 (46 Tau) Mass-Period Companion Phase Space for All Astrometry] 
{ \label{fig::26690_phase_space_all}
z-periodogram (left) and the mass-period companion phase space for HD 26690 (right).  
Companions in the regions above the plotted exclusion curve with circular 
orbits with any orientation are not consistent with the PHASES observations, 
with 99\% confidence.  Companions as small as 
9.6 Jupiter masses in stable orbits can be ruled out by the combined observations.}
\end{figure*}

HD 26690 (46 Tau, HR 1309, HIP 19719, WDS 04136$+$0743) is a single-lined 
spectroscopic binary with stellar components 
having masses near that of the Sun.  
The z-periodogram and mass-period space limits 
to tertiary companions using just 
the PHASES observations are presented in 
Figure \ref{fig::26690_phase_space_phases}.  
The FAP of the highest peak ($z = 9.83$ at $P=6.34$ days) in the z-periodogram 
is 2.8\%, with 99\% confidence at the $z=11.09$ level.  
Because the FAP is low, the search was re-evaluated using both the PHASES 
measurements as well as those from the {\em Washington Double Star Catalog} 
\citep[WDS, see references therein][]{wdsCatalog, wdsCatalogUpdate} 
as evaluated in Paper II.  Though lower precision, these 
measurements have better time coverage and thus help to avoid confusion with 
the motion of the binary itself.  The same set of companion orbital periods 
was selected for evaluation as for the PHASES-only search, since only the 
PHASES data are well-suited to identifying tertiary companions.  The synthetic 
data sets generated included synthetic measurements for the non-PHASES data as 
well; in these cases, Gaussian random values 
were selected in separation and position 
angle, with variances equal to the measurement uncertainties in the real data.
In this refined search, 
the highest peak has a value of only $z=3.28$ (at $P=6.31$ days), with 
an FAP of 67.1\% and 99\% confidence at $z=7.08$.  
Thus, it appears that the initial search did not identify a real 
companion.  The resulting z-periodogram and mass-period companion limits are 
presented in Figure \ref{fig::26690_phase_space_all}.

\subsection{HD 44926}

\begin{figure*}[!ht]
\plottwo{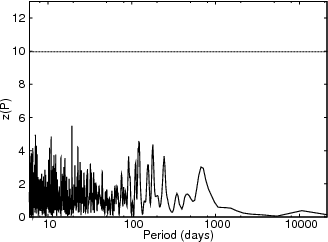}{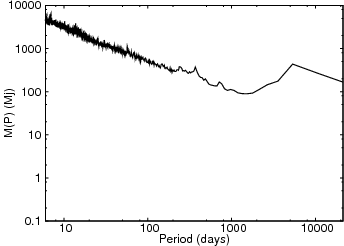}
\caption[HD 44926 Mass-Period Companion Phase Space] 
{ \label{fig::44926_phase_space}
z-periodogram (left) and the mass-period companion phase space for HD 44926 (right).  
Companions in the regions above the plotted exclusion curve with circular 
orbits with any orientation are not consistent with the PHASES observations, 
with 99\% confidence.  Companions as small as 
89 Jupiter masses can be ruled out by PHASES observations.}
\end{figure*}

HD 44926 (HIP 30569, WDS 06255$+$2327) is a 
relatively unstudied binary comprised 
of a pair of K giants.  The orbit and 
component masses are relatively uncertain, 
and the values listed here for the masses of companions that can be excluded 
are equally uncertain.  The FAP of the highest peak 
($z = 5.49$ at $P=19.3$ days) in the 
z-periodogram is 33.5\% and 99\% confidence 
would be found at $z=9.97$.  The 
z-periodogram and mass limits are plotted in Figure 
\ref{fig::44926_phase_space}.

\vspace{0.5in}
\subsection{HD 76943}
\begin{figure*}[!ht]
\plottwo{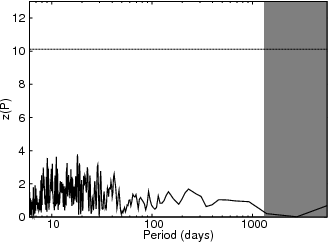}{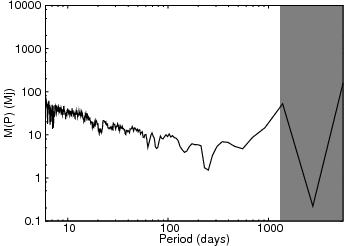}
\caption[HD 76943 (10 UMa) Mass-Period Companion Phase Space] 
{ \label{fig::76943_phase_space}
z-periodogram (left) and the mass-period companion phase space for HD 76943 (right).  
Companions in the regions above the plotted exclusion curve with circular 
orbits with any orientation are not consistent with the PHASES observations, 
with 99\% confidence.  Companions as small as 
1.5 Jupiter masses in stable orbits can be ruled out by PHASES observations.}
\end{figure*}

HD 76943 (10 UMa--though it is now in the constellation Lynx 
\citep{Griffin1999}, HR 3579, HIP 44248, WDS 09006$+$4147) is a relatively 
nearby double lined spectroscopic binary.  
The masses and system distance obtained 
by combining astrometry with velocities 
from TSU's AST in Paper II are not consistent 
with results from {\em Hipparcos} or the 
spectral types.  Thus, component masses and 
distance were used based on the {\em Hipparcos} results in \cite{Soder1999}.  
The FAP of the highest peak ($z = 3.78$ at $P=18.1$ days) in the z-periodogram 
is 62.9\% with 99\% detection confidence requiring $z=10.12$.  The periodogram 
and companion mass limits are plotted in Figure \ref{fig::76943_phase_space}.

\subsection{HD 77327}

\begin{figure*}[!ht]
\plottwo{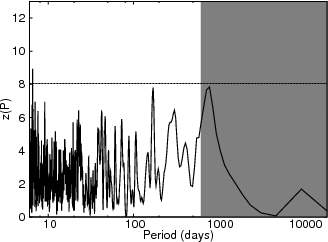}{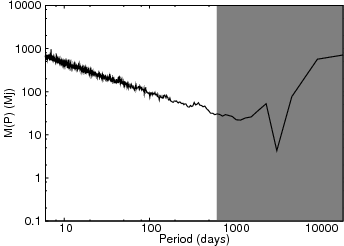}
\caption[HD 77327 (12 UMa) Mass-Period Companion Phase Space for PHASES Astrometry] 
{ \label{fig::77327_phase_space_phases}
z-periodogram (left) and the mass-period companion phase space for HD 77327 (right).  
Companions in the regions above the plotted exclusion curve with circular 
orbits with any orientation are not consistent with the PHASES observations, 
with 99\% confidence.  Companions as small as 
29 Jupiter masses in stable orbits can be ruled out by PHASES observations.}
\end{figure*}

\begin{figure*}[!ht]
\plottwo{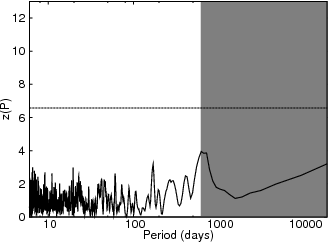}{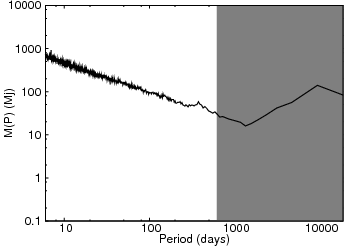}
\caption[HD 77327 (12 UMa) Mass-Period Companion Phase Space for All Astrometry] 
{ \label{fig::77327_phase_space_all}
z-periodogram (left) and the mass-period companion phase space for HD 77327 (right).  
Companions in the regions above the plotted exclusion curve with circular 
orbits with any orientation are not consistent with the PHASES observations, 
with 99\% confidence.  Companions as small as 32 Jupiter masses 
in stable orbits can be ruled out by the combined observations.}
\end{figure*}

HD 77327 ($\kappa$ UMa, 12 UMa, HR 3594, HIP 44471, WDS 09036$+$4709) 
is a pair of early A dwarf stars.  The total mass of the binary is only 
poorly constrained, so the values of companion masses ruled out by 
PHASES astrometry should be interpreted with a similar level of uncertainty.  
The FAP of the highest peak ($z = 8.94$ at $P=6.55$ days) in the z-periodogram 
is 0.3\%, and the 99\% confidence level for detection is at $z = 8.06$.  
This low FAP value prompted a second search, this time using 
both the PHASES and non-PHASES measurements, evaluated at the same perturbation 
orbital periods as in the PHASES-only search.  The addition of non-PHASES 
measurements helped define the long-term binary orbit, lifting fit 
degeneracies, 
and better identifying whether a detected perturbation was due to cadence and 
the wide orbit, or was evidence of 
a real companion.  The same procedure was used as 
for HD 26690.  The combined search showed a peak value of $z = 3.95$ with 
an FAP of 44.2\% and 99\% confidence of detection at $z=6.57$.  
Thus, it would appear that this was in fact 
a spurious detection, despite the low FAP.  
The z-periodograms and mass limits are 
plotted in Figures \ref{fig::77327_phase_space_phases} and 
\ref{fig::77327_phase_space_all}.

\subsection{HD 81858}

\begin{figure*}[!ht]
\plottwo{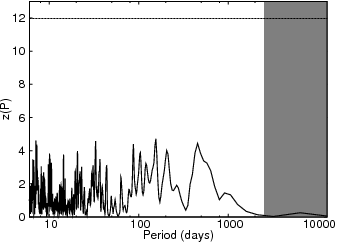}{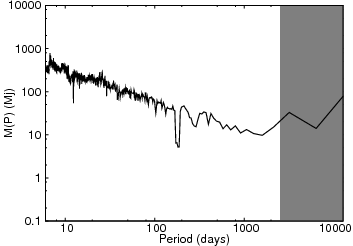}
\caption[HD 81858 ($\omega$ Leo) Mass-Period Companion Phase Space] 
{ \label{fig::81858_phase_space}
z-periodogram (left) and the mass-period companion phase space for HD 81858 (right).  
Companions in the regions above the plotted exclusion curve with circular 
orbits with any orientation are not consistent with the PHASES observations, 
with 99\% confidence.  Companions as small as 
5.2 Jupiter masses can be ruled out by PHASES observations.}
\end{figure*}

HD 81858 ($\omega$ Leo, 2 Leo, HR 3754, HIP 46454, WDS 09285$+$0903) is a single-lined 
spectroscopic binary.  The mass ratio is only poorly constrained by the available 
radial velocity data and parallax.  Thus, the average component mass of 
$1.10 \, \Msun$ was used to convert between astrometric perturbation amplitude and 
companion mass.  The FAP of the highest peak 
($z = 4.72$ at $P=155$ days) in the z-periodogram is 37.8\% with 99\% detection 
confidence only for signals with $z > 11.96$.  The resulting periodogram and 
mass-period space limits are presented in Figure \ref{fig::81858_phase_space}.

\subsection{HD 114378}

\begin{figure*}[!ht]
\plottwo{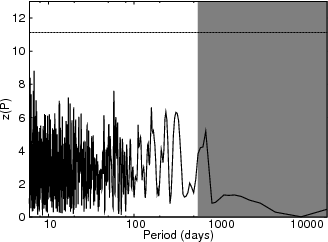}{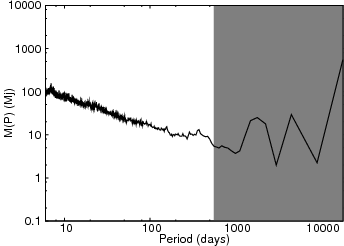}
\caption[HD 114378 ($\alpha$ Com) Mass-Period Companion Phase Space] 
{ \label{fig::114378_phase_space}
z-periodogram (left) and the mass-period companion phase space for HD 114378 (right).  
Companions in the regions above the plotted exclusion curve with circular 
orbits with any orientation are not consistent with the PHASES observations, 
with 99\% confidence.  Companions as small as 
6.3 Jupiter masses can be ruled out by PHASES observations.}
\end{figure*}

HD 114378 ($\alpha$ Com, 42 Com, HR 4968, HIP 64241, WDS 13100$+$1732) 
is a well studied long period binary.  It was included as a specific example 
system by the tertiary companion stability study of \cite{holman1999}.  The 
FAP of the highest peak ($z = 8.81$ at $P=6.81$ days) in the z-periodogram 
is 4.4\%, $z=11.13$ would be required for a reliable detection.  The 
periodogram and companion limits for HD 114378 are plotted 
in Figure \ref{fig::114378_phase_space}.

\subsection{HD 129246}

\begin{figure*}[!ht]
\plottwo{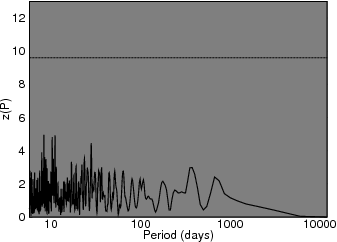}{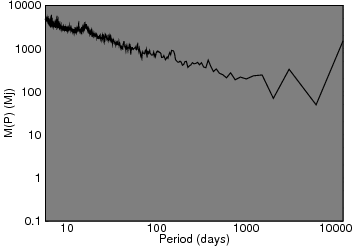}
\caption[HD 129246 ($\zeta$ Boo) Mass-Period Companion Phase Space] 
{ \label{fig::129246_phase_space}
z-periodogram (left) and the mass-period companion phase space for HD 129246 (right).  
Companions in the regions above the plotted exclusion curve with circular 
orbits with any orientation are not consistent with the PHASES observations, 
with 99\% confidence.}
\end{figure*}

HD 129246 ($\zeta$ Boo, 30 Boo, HR 5477, HIP 71795, WDS 1411$+$1344) 
has an extremely high eccentricity of $0.9977 \pm 0.0034$.  The distance of 
closest approach is only 0.3 AU.  It is unlikely any companions could 
have stable orbits in such a system.  The binary's eccentricity falls 
outside the regime examined by \cite{holman1999}, so it is not surprising 
their model breaks down in this regime.  The binary is useful as a test of 
the detection algorithm.  The FAP of the highest peak ($z = 4.95$ at $P=8.71$ 
days) in the z-periodogram is 25.4\% with a 1\% FAP occurring only for 
$z > 9.60$.  The periodogram and mass limits are plotted in 
Figure \ref{fig::129246_phase_space}.

\subsection{HD 137107}

\begin{figure*}[!ht]
\plottwo{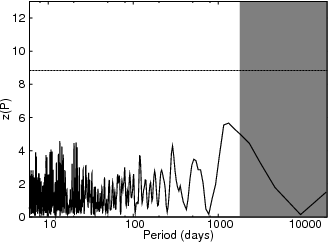}{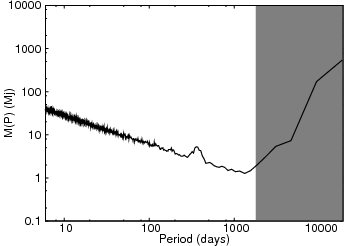}
\caption[HD 137107 ($\eta$ CrB) Mass-Period Companion Phase Space] 
{ \label{fig::137107_phase_space}
z-periodogram (left) and the mass-period companion phase space for HD 137107 (right).  
Companions in the regions above the plotted exclusion curve with circular 
orbits with any orientation are not consistent with the PHASES observations, 
with 99\% confidence.  Companions as small as 
1.3 Jupiter masses can be ruled out by PHASES observations.}
\end{figure*}

HD 137107 ($\eta$ CrB, 2 CrB, HR 5727, HIP 75312, WDS 15232$+$3017) 
is a double-lined 
spectroscopic binary comprised of stars just slightly more massive 
than the Sun.  It also has a distant (3600 AU), faint, brown dwarf companion 
in a circumbinary orbit \citep[which has no impact on the astrometric study 
of the A-B pair; ][]{2001AJ....121.3235K}.  The 
FAP of the highest peak ($z = 5.66$ at $P=1323$ days) in the 
z-periodogram is 19.9\% 
with a 1\% FAP occurring only for $z > 8.83$.  The periodogram and 
companion mass limits are plotted in Figure \ref{fig::137107_phase_space}.

\subsection{HD 137391}

\begin{figure*}[!ht]
\plottwo{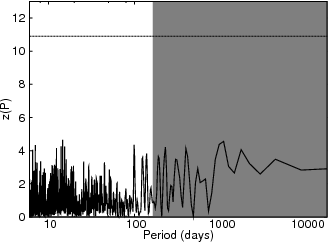}{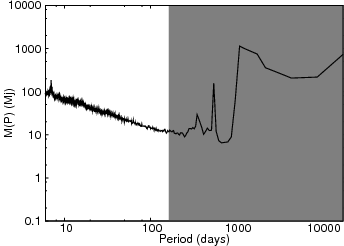}
\caption[HD 137391 ($\mu$ Boo) Mass-Period Companion Phase Space] 
{ \label{fig::137391_phase_space}
z-periodogram (left) and the mass-period companion phase space for HD 137391 (right).  
Companions in the regions above the plotted exclusion curve with circular 
orbits with any orientation are not consistent with the PHASES observations, 
with 99\% confidence.  Companions as small as 
11 Jupiter masses can be ruled out by PHASES observations.}
\end{figure*}

The periodogram and mass-period limits for HD 137391 
($\mu$ Boo, 51 Boo, HR 5733, HIP 75411, WDS 15245$+$3723) are plotted in Figure 
\ref{fig::137391_phase_space}.  The FAP of the highest peak ($z = 4.65$ at 
$P=14.6$ days) in the z-periodogram is 65.1\% with 99\% detection confidence 
requiring $z = 10.91$.

\subsection{HD 137909}
The primary of HD 137909 
(``Peculiar Rosette Stone'', $\beta$ CrB, 3 CrB, HR 5747, HIP 75695, WDS 15278$+$2906) is a 
prototype of the peculiar A stars along 
with $\gamma$ Equulei and $\alpha^2$ CVn .  
Given the increased frequency with which 
planets seem to occur around higher mass stars \citep{johnjohn2007} and those 
showing higher metallicities 
\citep{1997MNRAS.285..403G, 2004A&A...415.1153S, fischerValenti2005}, this is 
a particularly compelling target.  Furthermore, \cite{Neubauer1944} 
identified a second period of nearly a year ($P_2 \sim 320$ days) in radial 
velocity observations.  \cite{Kamper1990} presented new data that were 
inconsistent with the proposed perturbation, suggesting the orbital inclination 
had rotated to be face-on since the first half of that century.  The 
calculations based on \cite{holman1999} predict such a companion would not 
have a stable orbit.  Finally 
\cite{Soder1999} used {\em Hipparcos} astrometry to show no such companion 
could exist, a result verified by early PHASES results \citep{Mute06Limits}.

\begin{figure*}[!ht]
\plottwo{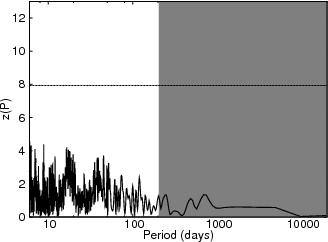}{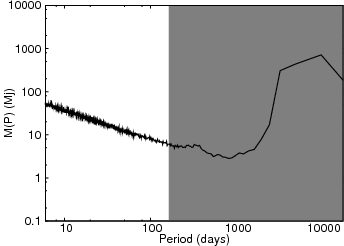}
\caption[HD 137909 ($\beta$ CrB) Mass-Period Companion Phase Space] 
{ \label{fig::137909_phase_space}
z-periodogram (left) and the mass-period companion phase space for HD 137909 (right).  
Companions in the regions above the plotted exclusion curve with circular 
orbits with any orientation are not consistent with the PHASES observations, 
with 99\% confidence.  Companions as small as 
4.8 Jupiter masses in stable orbits can be ruled out by PHASES observations.}
\end{figure*}

With the full PHASES data set being analyzed using the revised approach 
described in this paper, the FAP of the highest peak 
($z = 4.55$ at $P=6.07$ days) in the z-periodogram is 53.9\% with 99\% 
detection confidence requiring $z = 7.94$.  The periodogram 
and companion mass limits are plotted in Figure \ref{fig::137909_phase_space}.

\subsection{HD 140159}

\begin{figure*}[!ht]
\plottwo{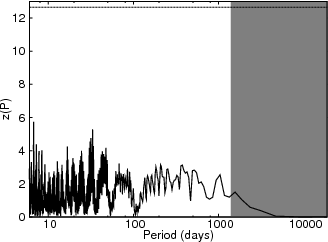}{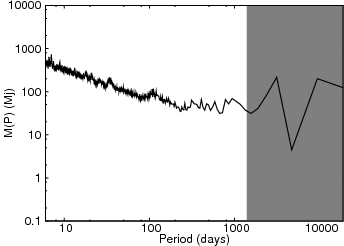}
\caption[HD 140159 ($\iota$ Ser) Mass-Period Companion Phase Space] 
{ \label{fig::140159_phase_space}
z-periodogram (left) and the mass-period companion phase space for HD 140159 (right).  
Companions in the regions above the plotted exclusion curve with circular 
orbits with any orientation are not consistent with the PHASES observations, 
with 99\% confidence.  Companions as small as 
31 Jupiter masses can be ruled out by PHASES observations.}
\end{figure*}

HD 140159 ($\iota$ Ser, 21 Set, HR 5842, HIP 76852, WDS 15416$+$1940) 
is a pair of early A dwarfs.  Being relatively massive stars a fairly large 
distance away, limits can only be placed on the existence of tertiary 
companions with masses in the brown dwarf or larger regime.  
The FAP of the highest peak ($z = 5.74$ at 
$P=6.72$ days) in the z-periodogram is 36.1\% with 99\% detection confidence 
requiring $z = 12.65$.  The periodogram 
and companion mass limits are plotted in Figure \ref{fig::140159_phase_space}.

\subsection{HD 140436}
\begin{figure*}[!ht]
\plottwo{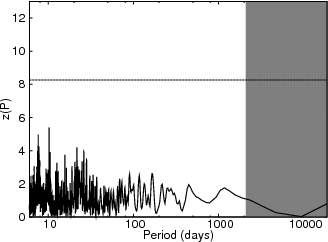}{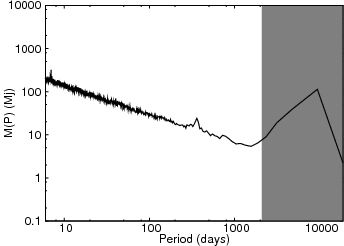}
\caption[HD 140436 ($\gamma$ CrB) Mass-Period Companion Phase Space] 
{ \label{fig::140436_phase_space}
z-periodogram (left) and the mass-period companion phase space for HD 140436 (right).  
Companions in the regions above the plotted exclusion curve with circular 
orbits with any orientation are not consistent with the PHASES observations, 
with 99\% confidence.  Companions as small as 
5.4 Jupiter masses can be ruled out by PHASES observations.}
\end{figure*}

Like HD 140159, HD 140436 
($\gamma$ CrB, 8 Crb, HR 5849, HIP 76952, WDS 15427$+$2618) is a pair of 
early A stars.  Both its binarity and early spectral type limit 
its ability to be studied by the radial velocity method for exoplanet searches, 
highlighting another manner in which astrometry can complement other 
techniques.  Objects as small as the largest of giant planets can be ruled out 
for some stable orbital periods in this system, despite the relatively 
large masses of the stars and distance to the system.  Some lower mass objects 
could have been detected if in fortunate orbital configurations (face-on 
orbits, or aligned parallel to the interferometer baseline vector)---as is 
the case for the other systems, the limits 
presented in Figure \ref{fig::140436_phase_space} consider all possible 
low-eccentricity orbits.  The FAP of the highest peak ($z = 5.40$ at 
$P=10.2$ days) in the z-periodogram is 23.8\% with 99\% detection confidence 
requiring $z = 8.26$.

\subsection{HD 155103}

\begin{figure*}[!ht]
\plottwo{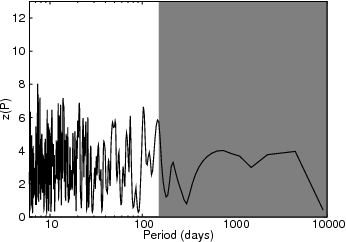}{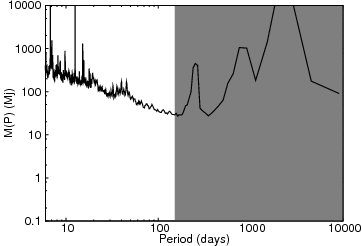}
\caption[HD 155103 (c Her) Mass-Period Companion Phase Space] 
{ \label{fig::155103_phase_space}
z-periodogram (left) and the mass-period companion phase space for HD 155103 (right).  
Companions in the regions above the plotted exclusion curve with circular 
orbits with any orientation are not consistent with the PHASES observations, 
with 99\% confidence.  Companions as small as 
29 Jupiter masses can be ruled out by PHASES observations.}
\end{figure*}

With only 10 PHASES measurements spanning just over two years, observing 
cadence causes more significant problems for HD 155103 
(c Her, HR 6377, HIP 83838, WDS 17080$+$3556) than most of the other 
binaries.  
The cadence results in multiple spikes in the mass-period sensitivity plot 
corresponding to orbital periods when orbit aliasing is more likely.  Combined 
with the relatively large mass of the components (1.66 $\Msun$) and distance 
to the system ($\sim 56$ pc), only limited constraints can be placed on 
tertiary companions.  The FAP of the highest peak ($z = 8.03$ at $P=7.36$ 
days) in the z-periodogram is 41.5\%, whereas 99\% confidence of detection 
would have only occurred for values larger than $z = 32.0$.  The periodogram 
and companion mass limits are plotted in Figure \ref{fig::155103_phase_space}.

\subsection{HD 187362}

\begin{figure*}[!ht]
\plottwo{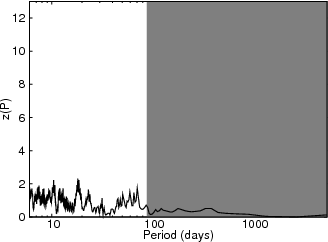}{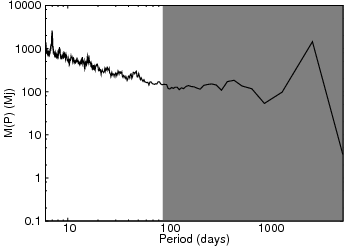}
\caption[HD 187362 ($\zeta$ Sge) Mass-Period Companion Phase Space] 
{ \label{fig::187362_phase_space}
z-periodogram (left) and the mass-period companion phase space for HD 187362 (right).  
Companions in the regions above the plotted exclusion curve with circular 
orbits with any orientation are not consistent with the PHASES observations, 
with 99\% confidence.  Companions as small as 
142 Jupiter masses can be ruled out by PHASES observations, roughly twice 
as massive as the largest of brown dwarfs.}
\end{figure*}

Like HD 155103, only 10 PHASES measurements of HD 187362 
($\zeta$ Sge, 8 Sge, HR 7546, HIP 97496, WDS 19490$+$1909) were made, though 
in this case with an even shorter timespan of 1.2 yr.  These relatively 
faint systems could not be observed until the instrument improvements were 
made that allowed the slower 20 Hz fringe tracking.  It too is relatively 
massive (average stellar mass 2.35 $\Msun$) and yet more distant 
($\sim 100$ pc) and only stellar mass objects in stable orbits can be 
excluded.  Also like HD 155103, the detection limit graph shows a jagged 
transition between the regions in which companions can and cannot be ruled out, 
due to observing cadence.  The FAP of the highest peak 
($z = 2.32$ at $P=18.1$ days) in the z-periodogram is 90.9\%, with 1\% FAP 
requiring $z=18.7$.  The periodogram 
and companion mass limits are plotted in Figure \ref{fig::187362_phase_space}.

\subsection{HD 202275}

\begin{figure*}[!ht]
\plottwo{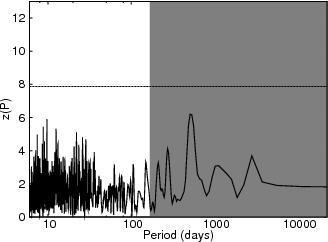}{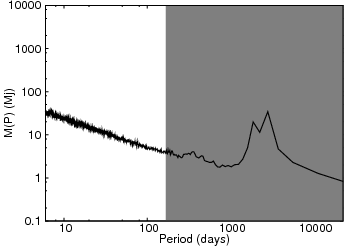}
\caption[HD 202275 ($\delta$ Equ) Mass-Period Companion Phase Space] 
{ \label{fig::202275_phase_space}
z-periodogram (left) and the mass-period companion phase space for HD 202275 (right).  
Companions in the regions above the plotted exclusion curve with circular 
orbits with any orientation are not consistent with the PHASES observations, 
with 99\% confidence.  Companions as small as 
3.8 Jupiter masses in stable orbits can be ruled out by PHASES observations.}
\end{figure*}

HD 202275 ($\delta$ Equ, 7 Equ, HR 8123, HIP 104858, WDS 21145$+$1000) was 
studied extensively by PHASES \citep{Mut06_kappeg, MuteMuOri2008}, 
with a span of observations of 1866 days, 
covering nearly the full binary orbit (2084 days).  Companions as small as 
3.8 Jupiter masses can be ruled out in stable orbits having any orientation.  
The FAP of the highest peak ($z = 6.20$ at $P=509$ days) in the z-periodogram
is 9.4\%, with 1\% FAP requiring $z=7.86$.  The periodogram 
and companion mass limits are plotted in Figure \ref{fig::202275_phase_space}.

\subsection{HD 202444}

There is some indication that $\tau$ Cyg may have a substellar 
companion orbiting one of the two stars (see Paper V).  There are reasons to 
doubt the authenticity of this proposed companion, so the visual 
orbit obtained by modeling the system with only a single Keplerian model 
has been presented in Paper II in addition to the double Keplerian model 
presented in Paper V.  If real, the companion has a long orbital 
period.  When only the shorter timespan PHASES data were analyzed to search 
for tertiary companions, the signal was absorbed into that of the wider 
binary, so no compelling evidence for a companion was present.  
However, the continued large value of $\chi^2$ 
that resulted when the combined PHASES and non-PHASES astrometry set was 
analyzed prompted a second search for tertiary companions, this time using all 
the astrometric measurements.  The longer timespan non-PHASES astrometry 
measurements better constrained the binary orbit parameters, preventing them 
from taking incorrect values to absorb the 
motion caused by an intermediate period companion (shorter than the 
binary motion, but long compared to the timespan of PHASES measurements) and 
indicated the presence of a companion with mass corresponding to that of 
a giant planet.

Because the companion only presents itself when both PHASES and 
non-PHASES measurements are jointly analyzed, it is more uncertain that the 
object is real.  This contrasts with the other candidate objects listed in 
Paper V, which could be detected both when just the PHASES measurements were 
considered and in the combined analysis.  For this reason, HD 202444 has been 
included in the present analysis to demonstrate what other companions can be 
shown not to exist in the case that the detected companion is not real either.  

\begin{figure*}[!ht]
\plottwo{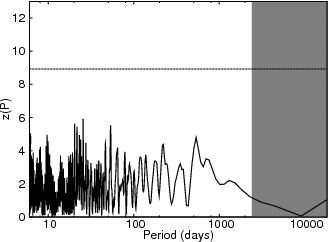}{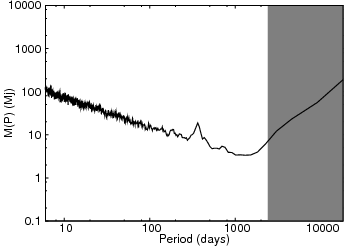}
\caption[HD 202444 ($\tau$ Cyg) Mass-Period Companion Phase Space for PHASES Astrometry] 
{ \label{fig::202444_phase_space_phases}
z-periodogram (left) and the mass-period companion phase space for HD 202444 (right), assuming that the candidate object is 
not real.  
Companions in the regions above the plotted exclusion curve with circular 
orbits with any orientation are not consistent with the PHASES observations, 
with 99\% confidence.  Companions as small as 
3.3 Jupiter masses in stable orbits can be ruled out by PHASES observations.}
\end{figure*}

\begin{figure*}[!ht]
\plottwo{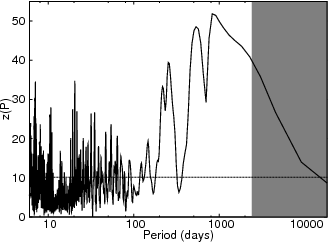}{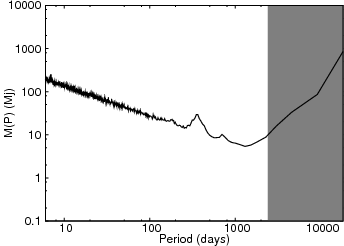}
\caption[HD 202444 ($\tau$ Cyg) Mass-Period Companion Phase Space for All Astrometry] 
{ \label{fig::202444_phase_space_all}
z-periodogram (left) and the mass-period companion phase space for HD 202444 (right), assuming that the candidate object is 
not real.
Companions in the regions above the plotted exclusion curve with circular 
orbits with any orientation are not consistent with the PHASES observations, 
with 99\% confidence.  Companions as small as 
5.4 Jupiter masses in stable orbits can be ruled out by the combined observations.}
\end{figure*}

For the PHASES-only analysis, the FAP of the highest peak 
($z = 5.93$ at $P=25.5$ days) in the z-periodogram is 19.1\%, 
with 1\% FAP requiring $z=8.92$.  However, when PHASES measurements are 
analyzed along with non-PHASES astrometry covering more of the binary orbit, 
the highest peak in the z-periodogram is $z = 51.9$ at $P=826$ days with an 
FAP of 0.0\%.  This peak is above the 1\% FAP mark, which would be at 
$z=10.1$.  The periodogram 
and companion mass limits when only PHASES observations are analyzed are 
plotted in Figure \ref{fig::202444_phase_space_phases} and those for the 
combined data set are plotted in Figure \ref{fig::202444_phase_space_all}, 
assuming the companion object is not real.  

\subsection{HD 207652}
\begin{figure*}[!ht]
\plottwo{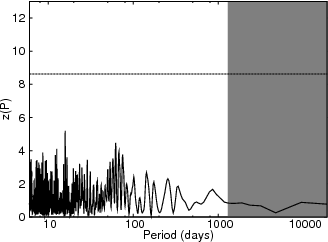}{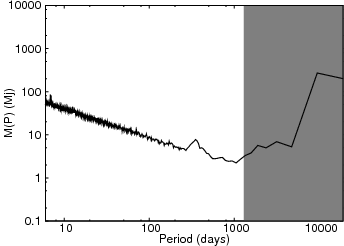}
\caption[HD 207652 (13 Peg) Mass-Period Companion Phase Space] 
{ \label{fig::207652_phase_space}
z-periodogram (left) and the mass-period companion phase space for HD 207652 (right).  
Companions in the regions above the plotted exclusion curve with circular 
orbits with any orientation are not consistent with the PHASES observations, 
with 99\% confidence.  Companions as small as 
2.2 Jupiter masses can be ruled out by PHASES observations.}
\end{figure*}

The periodogram and mass-period limits for HD 207652 
(13 Peg, HR 8344, HIP 107788, V373 Peg, WDS 21501$+$1717)
are plotted in Figure 
\ref{fig::207652_phase_space}.  The FAP of the highest peak ($z = 5.19$ at 
$P=15.8$ days) in the z-periodogram is 29.9\% with 99\% detection confidence 
requiring $z = 7.78$.  With a relatively large range of orbital periods that 
can be stable (up to 3.3 yr) and number of PHASES measurements (51), there 
is increased sensitivity to companion objects in HD 207652 than most of the 
other systems being considered.  Companions as small as 2.2 Jupiter masses can 
be ruled out in this binary.  It is worth noting that 
\cite{1999ApJ...513..933T} claim the secondary in the system is a T Tauri 
star, so this represents a possible non-detection of planets in a forming 
system.  

\subsection{HD 214850}

\begin{figure*}[!ht]
\plottwo{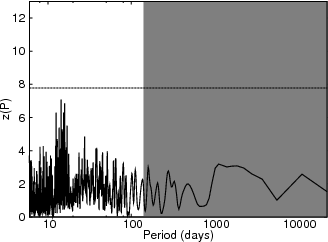}{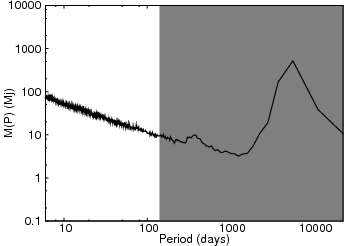}
\caption[HD 214850 (HR 8631) Mass-Period Companion Phase Space for PHASES Astrometry] 
{ \label{fig::214850_phase_space_phases}
z-periodogram (left) and the mass-period companion phase space for HD 214850 (right).  
Companions in the regions above the plotted exclusion curve with circular 
orbits with any orientation are not consistent with the PHASES observations, 
with 99\% confidence.  Companions as small as 
8.8 Jupiter masses in stable orbits can be ruled out by PHASES observations.}
\end{figure*}

\begin{figure*}[!ht]
\plottwo{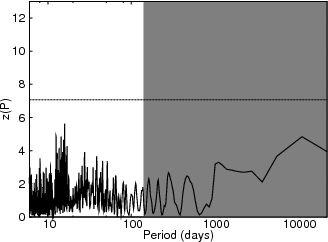}{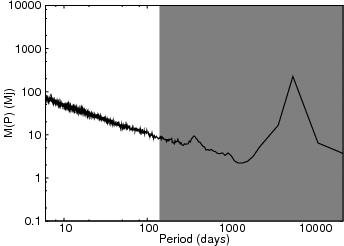}
\caption[HD 214850 (HR 8631) Mass-Period Companion Phase Space for All Astrometry] 
{ \label{fig::214850_phase_space_all}
z-periodogram (left) and the mass-period companion phase space for HD 214850 (right).  
Companions in the regions above the plotted exclusion curve with circular 
orbits with any orientation are not consistent with the PHASES observations, 
with 99\% confidence.  Companions as small as 
8.3 Jupiter masses in stable orbits can be ruled out by the combined observations.}
\end{figure*}

In the analysis of just the PHASES observations of HD 214850 
(HR 8631, HIP 111974, WDS 22409$+$1433), the FAP of the highest peak 
($z = 7.08$ at $P=14.4$ days) in the z-periodogram is 3.8\%.  The peak value 
$z=7.08$ is close to the 1\% FAP limit at $z=7.78$.  
As were the cases for HD 26690 and HD 77327, this low value 
inspired a second search including non-PHASES data 
from the WDS, as listed in Paper II.  
This revised search found a peak of $z = 5.63$ at $P=15.9$ days with an FAP of 
9.5\% and well below $z=7.06$ which would correspond to 1\% FAP .  Because the 
identified orbital period is different and the combined 
FAP is well beyond the 1\% threshold, there is not sufficient evidence to 
claim the existence of a companion object in this system.  The z-periodograms 
and mass-period phase space plots for the analysis of HD 214850 are shown in 
Figures \ref{fig::214850_phase_space_phases} and 
\ref{fig::214850_phase_space_all}.

\section{Future Directions}

The PHASES program used the same interferometric astrometry concepts as will be 
used in the SIM-Lite Astrometric Observatory 
mission \citep{SIM, 2008PASP..120...38U}.  SIM-Lite will benefit 
from greater stability and sensitivity that operating in a space environment 
allows, introducing improved measurement precisions and versatility.  
SIM-Lite astrometry operating on single stars can achieve measurement 
precisions over 1.5 orders of magnitude better than those presented 
here, with 10-100$\times \sim 30$ more measurements, on a much more flexible 
set of targets, including stars $\sim 10 \times$ closer to the solar system.  
Overall, this means a factor of $35 \times \sqrt{30} \times 10 \sim 2000$ 
better sensitivity to companions.  In addition, those measurements will be 
more two-dimensional than PHASES since the baseline will be rotated to two 
fully orthogonal directions.  SIM-Lite will move from the $\sim 10$ (typical) 
and $\sim 1$ (best) Jupiter-mass sensitivities of the present study into the 
regime of Earthlike planets.

\acknowledgements 
PHASES benefits from the efforts of the PTI collaboration members who have 
each contributed to the development of an extremely reliable observational 
instrument.  Without this outstanding engineering effort to produce a solid 
foundation, advanced phase-referencing techniques would not have been 
possible.  We thank PTI's night assistant Kevin Rykoski for his efforts to 
maintain PTI in excellent condition and operating PTI in phase-referencing 
mode every week.  Part of the work described in this paper was performed at 
the Jet Propulsion Laboratory under contract with the National Aeronautics 
and Space Administration.  Interferometer data were obtained at the Palomar
Observatory with the NASA Palomar Testbed Interferometer, supported
by NASA contracts to the Jet Propulsion Laboratory.  This publication makes 
use of data products from the Two Micron All Sky Survey, which is a joint 
project of the University of Massachusetts and the Infrared Processing and 
Analysis Center/California Institute of Technology, funded by the National 
Aeronautics and Space Administration and the National Science Foundation.  
This research has made use of the Simbad database, operated at CDS, 
Strasbourg, France.  M.W.M.~acknowledges support from the Townes Fellowship 
Program, Tennessee State University, and the state of Tennessee through its 
Centers of Excellence program.  Some of the software used for analysis was 
developed as part of the SIM Double Blind Test with support from NASA 
contract NAS7-03001 (JPL 1336910).  
PHASES is funded in part by the California Institute of Technology 
Astronomy Department, and by the National Aeronautics and Space Administration 
under grant no.~NNG05GJ58G issued through the Terrestrial Planet Finder 
Foundation Science Program.  This work was supported in part by the National 
Science Foundation through grants AST 0300096, AST 0507590, and AST 0505366.
M.K.~is supported by the Foundation for Polish Science through a FOCUS 
grant and fellowship, by the Polish Ministry of Science and Higher 
Education through grant N203 3020 35.

{\it Facilities:} \facility{PO:PTI}

\bibliography{main}
\bibliographystyle{apj}

\end{document}